\documentclass[aps,pre,twocolumn
]{revtex4-1}

\usepackage{mathrsfs}
\usepackage{bm}
\usepackage{amsmath}
\usepackage{amssymb}
\usepackage{graphicx}
\usepackage{times}
\usepackage{MnSymbol}
\usepackage{algorithm}
\usepackage{algpseudocode}

\makeatletter

\usepackage{hyperref}
\usepackage{color}
\definecolor{myblue}{RGB}{0,50,200}
\hypersetup{
colorlinks,
citecolor=myblue,
linkcolor=myblue,
urlcolor=myblue
}

\allowdisplaybreaks

\newcommand{\mca}{\mathcal}
\newcommand{\mbb}{\mathbb}

\newcommand{\msf}{\mathsf}
\newcommand{\avg}[1]{\langle #1\rangle}

\newcommand{\bra}[1]{\left( #1 \right)}
\newcommand{\bras}[1]{\left[ #1 \right]}
\newcommand{\brab}[1]{\left\{ #1 \right\}}
\newcommand{\pp}{\partial}
\newcommand{\Var}[1]{\llangle #1\rrangle}

\usepackage{mathtools}

\DeclarePairedDelimiter\floor{\lfloor}{\rfloor}

\begin{document}
\title{Entropy production estimation with optimal current}

\author{Tan Van Vu}
\email{tan@biom.t.u-tokyo.ac.jp}

\affiliation{Department of Information and Communication Engineering, Graduate
School of Information Science and Technology, The University of Tokyo,
Tokyo 113-8656, Japan}

\author{Van Tuan Vo}
\email{tuan@biom.t.u-tokyo.ac.jp}

\affiliation{Department of Information and Communication Engineering, Graduate
School of Information Science and Technology, The University of Tokyo,
Tokyo 113-8656, Japan}

\author{Yoshihiko Hasegawa}
\email{hasegawa@biom.t.u-tokyo.ac.jp}

\affiliation{Department of Information and Communication Engineering, Graduate
School of Information Science and Technology, The University of Tokyo,
Tokyo 113-8656, Japan}

\date{\today}

\begin{abstract}
Entropy production characterizes the thermodynamic irreversibility and reflects the amount of heat dissipated into the environment and free energy lost in nonequilibrium systems.
According to the thermodynamic uncertainty relation, we propose a deterministic method to estimate the entropy production from a single trajectory of system states.
We explicitly and approximately compute an optimal current that yields the tightest lower bound using predetermined basis currents.
Notably, the obtained tightest lower bound is intimately related to the multidimensional thermodynamic uncertainty relation.
By proving the saturation of the thermodynamic uncertainty relation in the short-time limit, the exact estimate of the entropy production can be obtained for overdamped Langevin systems, irrespective of the underlying dynamics.
For Markov jump processes, because the attainability of the thermodynamic uncertainty relation is not theoretically ensured, the proposed method provides the tightest lower bound for the entropy production.
When entropy production is the optimal current, a more accurate estimate can be further obtained using the integral fluctuation theorem.
We illustrate the proposed method using three systems: a four-state Markov chain, a periodically driven particle, and a multiple bead-spring model.
The estimated results in all examples empirically verify the effectiveness and efficiency of the proposed method.
\end{abstract}

\maketitle

\section{Introduction}
Entropy production is a fundamental thermodynamic quantity that characterizes the irreversibility of thermodynamic processes.
Owing to the development of stochastic thermodynamics \cite{Sekimoto.1998.PTPS,Seifert.2012.RPP}, a mesoscopic expression of entropy production has been formulated in the trajectory level \cite{Seifert.2005.PRL,Spinney.2012.PRL}.
As a consequence, a universal property regarding the symmetry of the probability distribution of entropy production was discovered as the fluctuation theorem \cite{Evans.1993.PRL,Gallavotti.1995.PRL,Crooks.1999.PRE}, from which the second law of thermodynamics can be derived.
Entropy production quantifies dissipation costs in nonequilibrium systems and is essential in the fundamental limits of the efficiency of physical systems, such as heat engines and refrigerators \cite{Salas.2010.PRE,Pietzonka.2018.PRL,Vroylandt.2019.JSM}.
In the context of biological processes, entropy production indicates the free energy lost in the spontaneous relaxation to perform a specific function \cite{Ge.2010.PRE}.
Therefore, the estimation of entropy production from the experimental data allows us to access the limits that cannot be exceeded and also provides insight into the underlying mechanism of physical systems \cite{Gnesotto.2018.RPP}.

Recent studies have made considerable advances in the entropy production inference based on the time-series data \cite{Lander.2012.PRE,Martinez.2019.NC,Li.2019.NC,Manikandan.2019.arxiv}.
Inference strategies can be generally classified into two classes: direct and indirect.
The authors in Ref.~\cite{Li.2019.NC} employed the former class to quantify dissipation for systems described by the additive-noise Langevin equations; the detailed dynamics of the system (e.g., drift terms and probability fluxes) were estimated, and the associated entropy production was subsequently approximated by either a spatial or a temporal average.
However, with an increase in the dimensionality, this strategy becomes computationally costly, and a prohibitive amount of data is required to accurately estimate the underlying dynamics.
Furthermore, the direct strategy is not applicable to situations wherein the full freedom degrees of the system cannot be observed in the experiments (e.g., some hidden variables exist due to the resolution limit of the measuring instrument \cite{Mehl.2012.PRL,Shiraishi.2015.PRE}).
Alternatively, an indirect strategy based on important recent discoveries called thermodynamic uncertainty relations (TURs) \cite{Barato.2015.PRL,Gingrich.2016.PRL,Horowitz.2017.PRE,Pietzonka.2016.PRE,Polettini.2016.PRE,Proesmans.2017.EPL,Garrahan.2017.PRE,Dechant.2018.JSM,Barato.2018.NJP,Macieszczak.2018.PRL,Brandner.2018.PRL,Hasegawa.2019.PRE,Koyuk.2019.PRL,Vu.2019.PRE.Underdamp,Chun.2019.PRE,Barato.2019.JSM,Vu.2019.PRE.Delay,Dechant.2019.JPA,Hasegawa.2019.PRL,Guarnieri.2019.PRR,Proesmans.2019.JSM,Pigolotti.2017.PRL,Vu.2020.JPA,Potts.2019.PRE,Falasco.2019.arxiv,Timpanaro.2019.PRL,Lee.2019.PRE,Vu.2020.PRR,Hasegawa.2019.arxiv.QuanTUR} (see \cite{Horowitz.2019.NP} for review), was proposed \cite{Li.2019.NC,Manikandan.2019.arxiv}.
TURs impose the following bound for steady-state systems described by continuous-time Markov jump processes and overdamped Langevin dynamics:
\begin{equation}
\Sigma\ge\frac{2\avg{\phi}^2}{\tau\Var{\phi}},\label{eq:org.TUR}
\end{equation}
where $\phi$ is an arbitrary time-integrated current, $\avg{\phi}$ and $\Var{\phi}:=\avg{\phi^2}-\avg{\phi}^2$ are its mean and variance, respectively, $\tau$ is the observation time, and $\Sigma$ is the entropy production rate.
Theoretically, a lower bound of entropy production can be obtained using a TUR.
Specifically, when the equality in Eq.~\eqref{eq:org.TUR} is attained, the exact entropy production inference is possible \cite{Manikandan.2019.arxiv}.
TUR appears to be a powerful tool for entropy production inference; however, an efficient method is still in development from the practical perspective.

In this study, we propose a deterministic method of entropy production estimation that is based on the TUR for classical Markovian dynamics.
We compute a current that maximizes the lower bound (i.e., minimizes its relative fluctuation) and is referred to as the optimal current.
For overdamped Langevin dynamics, we rigorously prove that TUR can be saturated in the short-time limit with the current of entropy production, even when the system is arbitrarily far from equilibrium.
Therefore, entropy production can be accurately estimated via the fluctuation of the optimal current in the short-time limit.
For Markov jump processes, we construct a counterexample in which TUR is unattainable with the current of entropy production.
Accordingly, entropy production is not guaranteed to be exactly estimated as in the case of Langevin dynamics.
In this case, our method provides the tightest possible lower bound on the entropy production.
However, given that entropy production is the optimal current, an exact estimate can be further obtained by combining our method with the fluctuation theorem.
We illustrate our approach with the help of three systems: a four-state Markov jump process, a periodically driven nonlinear system, and a tractable bead-spring model.
The results demonstrate that the proposed method produces accurate estimates of entropy production for Langevin systems, and the tightest lower bound for Markov jump processes.
Notably, the computed optimal current accurately approximates the stochastic entropy production, which agrees with the theory that the entropy production is one of the optimal currents in the Langevin dynamics.

Indirect inference on the basis of the TUR has several advantages over the direct one.
First, it can robustly estimate a lower bound on entropy production even in the presence of hidden variables, while the direct strategy cannot.
This situation is common in the biological context, where the full degrees of freedom are often inaccessible.
Second, for Langevin dynamics involving multiplicative noises, the accurate estimation of both the drift and diffusion terms is not a simple task, especially in the high-dimensional case.
Moreover, the errors that occurred in the estimation of these quantities can be accumulated in the phase of calculating entropy production, which potentially affects the accuracy of the estimate.
In contrast, inference that is based on the TUR does not require us to know the underlying dynamics, e.g., whether the noises are additive or multiplicative.

\section{Method}
\begin{figure}[t]
\centering
\includegraphics[width=\linewidth]{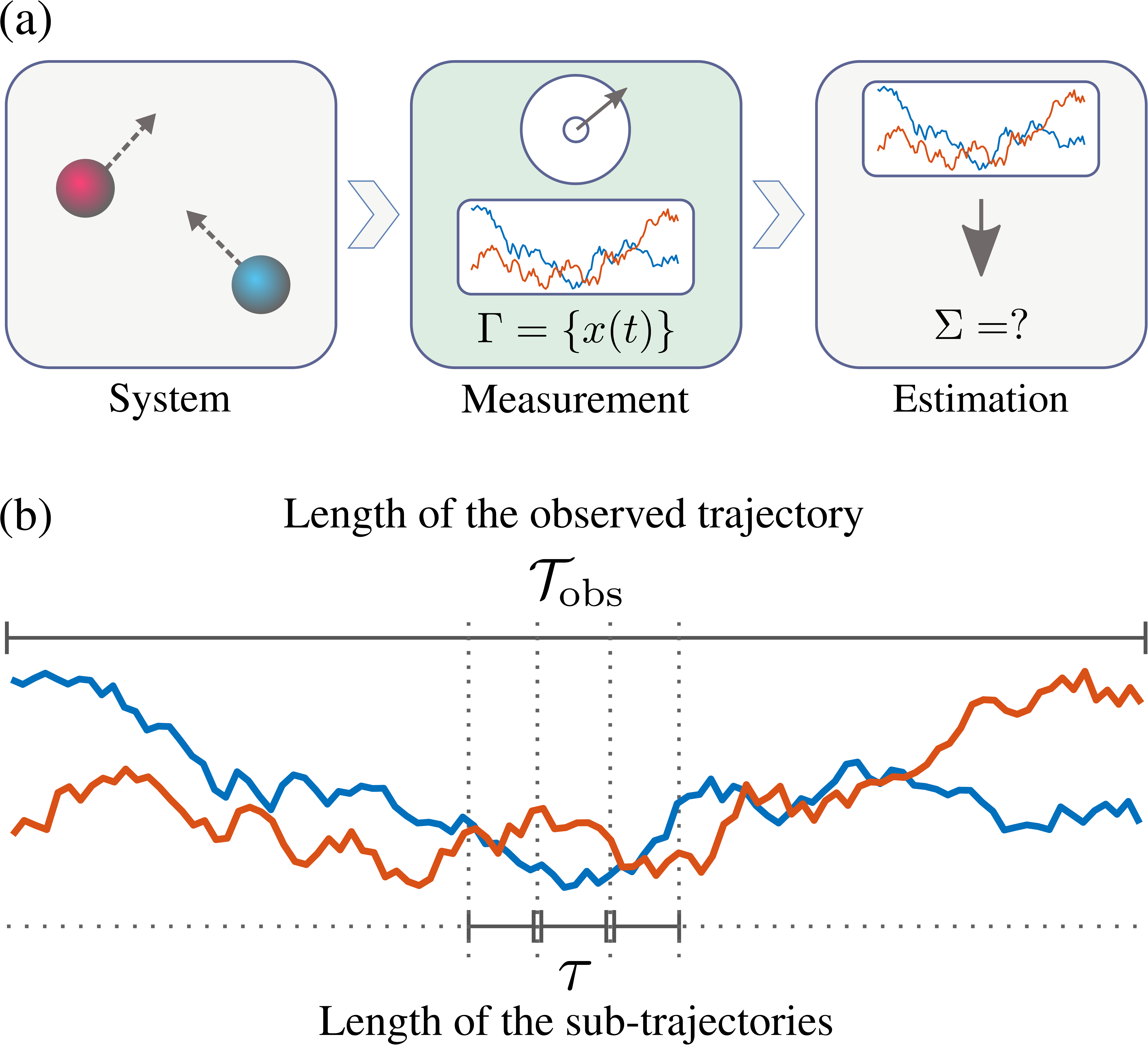}
\protect\caption{(a) Schematic diagram of entropy production estimation. A trajectory $\Gamma=\{x(t)\}_{t=0}^{t=\mca{T}_{\rm obs}}$ of the steady-state system is observed by a measuring instrument. Then, the entropy production rate $\Sigma$ is estimated solely from this single trajectory. (b) Schematic diagram of the trajectory-split process. The observed trajectory of length $\mca{T}_{\rm obs}$ is split into multiple sub-trajectories of length $\tau$ ($\ll\mca{T}_{\rm obs}$). Note that the sub-trajectories can be overlapped in the splitting phase to increase the number of samples.}\label{fig:Setup}
\end{figure}

In this section, we describe our method of entropy production estimation for both Markov jump processes and Langevin dynamics.
First, we discuss the strategy of entropy production estimation on the basis of TUR.
Then, we explain in detail how to efficiently estimate entropy production in practice.
The procedure of entropy production estimation is illustrated in Fig.~\ref{fig:Setup}.

\subsection{Entropy production estimation on the basis of TUR}
The lower bound of the entropy production rate can be estimated from TUR [Eq.~\eqref{eq:org.TUR}] as
\begin{equation}
\Sigma\ge\widehat{\Sigma}_\tau:=\max_{\phi}\frac{2\avg{\phi}^2}{\tau\Var{\phi}},\label{eq:est.formula.TUR}
\end{equation}
where the maximum is taken over all possible currents.
The inequality \eqref{eq:est.formula.TUR} immediately suggests a simple way to obtain the lower bound of the entropy production rate as follows: (i) observing a variety of currents in the system and calculating the fluctuation of each current and (ii) setting a maximum of $\{2\avg{\phi}^2/\tau\Var{\phi}\}$ as a lower bound on $\Sigma$.
Despite its simplicity, there are several issues when employing this strategy.
First, there is no theory that supports the number and the detailed forms of currents needed to yield a good estimate.
Moreover, it is also difficult to assess whether the present maximum value is the tightest bound or not.
Clearly, if the explicit form of the optimal current is known in advance, one can observe such a current and readily obtain the tightest bound for the entropy production rate.
Given the underlying dynamics, a recent study has proposed a method to analytically calculate the optimal current, which is called the hyper-accurate current \cite{Busiello.2019.PRE}.
Without accessibility to the details of dynamics, it is impossible to attain an exact form.
In Ref.~\cite{Li.2019.NC}, the authors used the Monte Carlo method to randomly sample the optimal current.
However, the resulting current is only sub-optimal when the system is strongly driven from equilibrium.
In the next section, we propose a deterministic strategy to efficiently approximate the optimal current from a single trajectory.

To obtain an exact estimate of the entropy production rate, the saturation in Eq.~\eqref{eq:est.formula.TUR} is required, i.e., $\widehat{\Sigma}_\tau=\Sigma$.
Recently, the authors in Ref.~\cite{Manikandan.2019.arxiv} stated that the equality can be attained in the short-time limit with the current $\sigma$ of entropy production, i.e.,
\begin{equation}
\frac{2\avg{\sigma}^2}{\tau\Var{\sigma}}\xrightarrow{\tau\to 0}\Sigma,~\text{or}~\mca{F}:=\frac{\Var{\sigma}}{\avg{\sigma}}\xrightarrow{\tau\to 0}2.\label{eq:opt.cur.ent}
\end{equation}
Here, we use the relation $\avg{\sigma}=\tau\Sigma$, and $\mca{F}$ denotes the Fano factor of $\sigma$.
Equation \eqref{eq:opt.cur.ent} implies that for short observation times, $\sigma$ is the optimal current, and its Fano factor $\mca{F}$ converges to $2$.
However, we show that this statement holds for overdamped Langevin dynamics but not for the Markov jump processes.
We rigorously prove that for systems described by overdamped Langevin equations, the Fano factor of entropy production always converges to $2$ in the short-time limit.
The details of the proof are presented in Appendix~\ref{app:Langevin}.
Regarding Markov jump processes, we construct a counterexample, in which $\mca{F}$ can be arbitrarily large even in the short-time limit.
The details of the counterexample are provided in Appendix~\ref{app:Markov.jump}.
In conclusion, the entropy production rate can be accurately estimated for Langevin dynamics.
However, only the tightest lower bound on the entropy production rate can be obtained for Markov jump processes.

\subsection{Approximation of the optimal current}
Let $\mca{C}=\{\phi_i(\Gamma)\}_{i=1}^n$ be a set of predetermined basis currents such that an arbitrary current can be approximately formed as a linear combination of these currents.
Here, $\Gamma$ denotes a given trajectory, and $n$ is the number of basis currents.
The construction of $\mca{C}$ (i.e., how to define the detailed form of each basis current $\phi_i$) will be described in the next section.
We assume that the optimal current can be expressed in terms of basis currents as $\phi_{\rm opt}(\Gamma)=\sum_{i=1}^{n}c_i\phi_i(\Gamma)$, where $\bm{c}=[c_1,\dots,c_n]^\top\in\mbb{R}^{n\times 1}$ is the coefficient vector.
Then, the mean and variance of $\phi_{\rm opt}$ can be analytically calculated via the basis currents as
\begin{align}
\avg{\phi_{\rm opt}}&=\bm{c}^\top\bm{\mu},\\
\Var{\phi_{\rm opt}}&=\bm{c}^\top\Xi\bm{c},
\end{align}
where $\bm{\mu}:=[\avg{\phi_1},\dots,\avg{\phi_n}]^\top\in\mbb{R}^{n\times 1}$ and $\Xi:=[\avg{\phi_i\phi_j}-\avg{\phi_i}\avg{\phi_j}]\in\mbb{R}^{n\times n}$ denote the means and the covariance matrix of the basis currents, respectively.
The computation of $\phi_{\rm opt}$ is equivalent to finding the optimal value of $\bm{c}$ that maximizes the following function:
\begin{equation}
\mca{J}(\bm{c})=\frac{\avg{\phi_{\rm opt}}^2}{\Var{\phi_{\rm opt}}}=\frac{\mca{E}(\bm{c})^2}{\mca{V}(\bm{c})},
\end{equation}
where $\mca{E}(\bm{c})=\bm{c}^\top\bm{\mu}$ and $\mca{V}(\bm{c})=\bm{c}^\top\Xi\bm{c}$.
Fortunately, this optimization problem can be solved analytically.
Since $\mca{J}(\bm{c})$ is scale-invariant with respect to $\bm{c}$, i.e., $\mca{J}(\kappa\bm{c})=\mca{J}(\bm{c})~\forall\kappa\neq 0$, we can add an equality constraint, $\mca{E}(\bm{c})=1$.
Consequently, the maximizing $\mca{J}(\bm{c})$ and minimizing $\mca{V}(\bm{c})$ optimizations are equivalent.
The latter optimization can be exactly solved using the Lagrange multipliers method.
We consider the Lagrangian function
\begin{equation}
\mca{L}(\bm{c},\lambda)=\frac{1}{2}\mca{V}(\bm{c})-\lambda(\mca{E}(\bm{c})-1).
\end{equation}
Taking the partial derivative of $\mca{L}$ with respect to $c_i~(i=1,\dots,n)$ and $\lambda$, we obtain
\begin{align}
0&=\pp_{c_i}\mca{L}(\bm{c},\lambda)=\sum_{j=1}^nc_j\Xi_{ij}-\lambda\mu_i,~(i=1,\dots,n),\label{eq:Lagrangian.c}\\
0&=\pp_{\lambda}\mca{L}(\bm{c},\lambda)=1-\sum_{i=1}^nc_i\mu_i.\label{eq:Lagrangian.lambda}
\end{align}
By solving Eqs.~\eqref{eq:Lagrangian.c} and \eqref{eq:Lagrangian.lambda}, the explicit solution is obtained
\begin{equation}
\lambda=(\bm{\mu}^\top\Xi^{-1}\bm{\mu})^{-1},~\bm{c}=(\bm{\mu}^\top\Xi^{-1}\bm{\mu})^{-1}\Xi^{-1}\bm{\mu}.\label{eq:opt.c.lambda}
\end{equation}
Thus, the maximum value of $\mca{J}(\bm{c})$ is
\begin{equation}
\mca{J}_{\rm max}:=\max_{\bm{c}}\mca{J}(\bm{c})=\bm{\mu}^\top\Xi^{-1}\bm{\mu}.
\end{equation}
Since the fluctuation of the optimal current $\phi_{\rm opt}$ obeys TUR, we have
\begin{equation}
\frac{2\avg{\phi_{\rm opt}}^2}{\tau\Var{\phi_{\rm opt}}}=\frac{2\mca{J}_{\rm max}}{\tau}=\frac{2\bm{\mu}^\top\Xi^{-1}\bm{\mu}}{\tau}\le\Sigma.\label{eq:tightest.bound}
\end{equation}
Equation~\eqref{eq:tightest.bound} implies that $\widehat{\Sigma}_\tau=2\mca{J}_{\rm max}/\tau$ is the tightest lower bound for the entropy production rate $\Sigma$ for the given set of basis currents $\mca{C}$.
Because TUR can be saturated in the short-time limit for Langevin dynamics, this lower bound is expected to be exactly the entropy production rate.
Moreover, as shown later, an arbitrary current in the Markov jump process can always be exactly expressed in the form of a linear combination of basis currents; thus, $2\mca{J}_{\rm max}/\tau$ is the tightest lower bound on the entropy production rate for arbitrary observation times.
Using the coefficient vector $\bm{c}$ obtained in Eq.~\eqref{eq:opt.c.lambda}, the optimal current can be readily calculated as $\phi_{\rm opt}=\sum_{i=1}^nc_i\phi_i$.
The obtained optimal current is in agreement with that reported in Ref.~\cite{Macieszczak.2018.PRL}.
The inequality $\bm{\mu}^\top\Xi^{-1}\bm{\mu}\le\tau\Sigma/2$ is also a consequence of the multidimensional TUR \cite{Dechant.2019.JPA,Vu.2019.PRE.Underdamp}, which provides a tighter bound than that of the scalar TUR [Eq.~\eqref{eq:org.TUR}].
Here, our analysis indicates that the multidimensional TUR has a remarkable application in the entropy production estimation, which was not revealed until now.

We summarize the procedure of estimating entropy production rate in the following.
\begin{algorithm}[H]
\caption{Estimation of the entropy production rate}
\label{algo:infer.ent}
\begin{algorithmic}[1]
\Require A given trajectory of system states $\Gamma=\{x(t)\}_{t=0}^{t=\mca{T}_{\rm obs}}$
\Ensure The estimated entropy production rate $\widehat{\Sigma}_\tau$
\State Define a set of basis currents $\mca{C}=\{\phi_1,\dots,\phi_n\}$
\State Split $\Gamma$ into sub-trajectories $\{\Gamma_k\}$ of length $\tau$ as [see Fig.~\ref{fig:Setup}(b)]
\State Compute $\mu_i=\avg{\phi_i},\Xi_{ij}=\avg{\phi_i\phi_j}-\avg{\phi_i}\avg{\phi_j}$ using $\{\Gamma_k\}$
\State Calculate optimal coefficients $\bm{c}=(\bm{\mu}^\top\Xi^{-1}\bm{\mu})^{-1}\Xi^{-1}\bm{\mu}$
\State Return $\widehat{\Sigma}_\tau=2\bm{\mu}^\top\Xi^{-1}\bm{\mu}/\tau$
\end{algorithmic}
\end{algorithm}
The statistical values of $\phi_i$ can be numerically approximated from sub-trajectories as
\begin{align}
\avg{\phi_i}&=\frac{1}{\mca{N}_\Gamma}\sum_k\phi_i(\Gamma_k),\\
\avg{\phi_i\phi_j}&=\frac{1}{\mca{N}_\Gamma}\sum_k\phi_i(\Gamma_k)\phi_j(\Gamma_k),
\end{align}
where $\mca{N}_\Gamma:=|\{\Gamma_k\}|$ denotes the number of sub-trajectories.

\subsection{Construction of basis currents}
Here, we describe the construction of basis currents for continuous-time Markov jump processes and overdamped Langevin dynamics.

\subsubsection{Markov jump process}

We consider a system modeled by the continuous-time Markov jump process on a finite countable state space $\Omega$.
Its dynamics are governed by the master equation
\begin{equation}
\pp_t p(y,t)=\sum_{z\in\Omega}\bras{p(z,t)w_{zy}-p(y,t)w_{yz}},
\end{equation}
where $p(y,t)$ denotes the probability distribution at time $t$ and $w_{yz}$ denotes the transition rate from state $y$ to state $z$.
We assume that $w_{zy}>0$ whenever $w_{yz}>0$, and the system always relaxes to a unique steady state in the long-time limit.
Let $p^{\rm ss}(y)$ denote the steady-state distribution, which satisfies $\sum_{z\in\Omega}\bras{w_{zy}p^{\rm ss}(z)-w_{yz}p^{\rm ss}(y)}=0,~\forall y\in\Omega$.

Given a trajectory $\Gamma=[x(t)]_{t=0}^\tau$, a generic current in the system can be represented as
\begin{equation}
\phi(\Gamma)=\sum_{y<z}\gamma_{yz}\int_0^\tau dt\,(\delta_{x(t^-),y}\delta_{x(t^+),z}-\delta_{x(t^-),z}\delta_{x(t^+),y}),
\end{equation}
where $\gamma_{yz}$'s are arbitrary real numbers, and $x(t^{-})$ and $x(t^{+})$ denote the state of the system immediately before and after a jump, respectively.
Define the set of basis currents as $\mca{C}=\{\phi_{yz}\}_{y<z}$, where $\phi_{yz}(\Gamma)=\int_0^\tau dt\,(\delta_{x(t^-),y}\delta_{x(t^+),z}-\delta_{x(t^-),z}\delta_{x(t^+),y})$ is a current that counts the net number of jumps between $y$ and $z$.
Then, the arbitrary current $\phi$ can be written in terms of basis currents $\{\phi_{yz}\}$ as
$\phi(\Gamma)=\sum_{y<z}\gamma_{yz}\phi_{yz}(\Gamma)$. 
For example, the current of stochastic entropy production has the form \cite{Seifert.2012.RPP}
\begin{equation}
\sigma(\Gamma)=\sum_{y<z}\ln\frac{p^{\rm ss}(y)w_{yz}}{p^{\rm ss}(z)w_{zy}}\phi_{yz}(\Gamma),
\end{equation}
which corresponds to the case $\gamma_{yz}=\ln p^{\rm ss}(y)w_{yz}/p^{\rm ss}(z)w_{zy}$.
Because arbitrary currents can always be expressed as a linear combination of basis currents $\{\phi_{yz}\}$, the optimal current $\phi_{\rm opt}$ can be accurately computed.
Equivalently, the tightest lower bound on the entropy production rate can always be obtained.

In special cases, the entropy production rate can, in principle, be accurately estimated using additional steps, even when the optimal current does not saturate the TUR.
If the entropy production is the optimal current, i.e., $\sigma=\alpha\phi_{\rm opt}$, where $\alpha$ is an unknown scaling factor, then $\Sigma$ can be estimated by employing the integral fluctuation theorem as follows.
First, $\alpha$ can be determined by examining whether the relation $\avg{e^{-\sigma}}=1$ holds or not.
Specifically, this is equivalent to solving the equation $\Psi(\alpha)=\mca{N}_\Gamma$, where $\Psi(\alpha)=\sum_{k}e^{-\alpha\phi_{\rm opt}(\Gamma_k)}$.
Here, we consider the case where the trajectory $\Gamma$ is well sampled; that is, both negative and positive values are contained in $\{\phi_{\rm opt}(\Gamma_k)\}_k$.
Since $\Psi(\alpha)$ is a convex function and $\Psi(0)=\mca{N}_\Gamma,~\Psi(-\infty)=\Psi(\infty)=\infty$, this equation has at most one nonzero solution, which can be, if it exists, efficiently computed using the Newton--Raphson method.
After obtaining $\alpha$, the entropy production rate can be readily estimated as $\widehat{\Sigma}_\tau=\alpha\avg{\phi_{\rm opt}}/\tau$.
It was proved that the entropy production is the optimal current for the long-time limit \cite{Gingrich.2016.PRL}.
However, the stochastic entropy production tends to be positive in this limit and the negative samples are rare.
Thus, the equation $\Psi(\alpha)=\mca{N}_\Gamma$ may have only the trivial solution $\alpha=0$, which means that the entropy production rate cannot be further estimated.

\subsubsection{Langevin dynamics}
For simplicity, we consider a one-dimensional system whose dynamics are described by the Langevin equation,
\begin{equation}
\dot{x}=F(x)+\sqrt{2D(x)}\xi(t),\label{eq:Ito.Lan.eq}
\end{equation}
where $F(x)$ is the force, $D(x)>0$ is the diffusion term, and $\xi$ is the zero-mean Gaussian white noise with a variance of $\avg{\xi(t)\xi(t')}=\delta(t-t')$.
The noise term in Eq.~\eqref{eq:Ito.Lan.eq}, $\sqrt{2D(x)}\xi$, is interpreted in the Ito sense.
Boltzmann's constant and the friction coefficient are set to unity throughout this study.
Let $p(x,t)$ denote the probability distribution function of the system state at time $t$.
Then, the corresponding Fokker--Planck equation is written as
\begin{equation}
\pp_tp(x,t)=-\pp_xj(x,t),
\end{equation}
where $j(x,t)=F(x)p(x,t)-\pp_x[D(x)p(x,t)]$ is the probability current.
Again, we focus exclusively on the steady state, where $p(x,t)=p^{\rm ss}(x)$ and $j(x,t)=j^{\rm ss}$.
The current of stochastic entropy production is expressed as \cite{Spinney.2012.PRE}
\begin{equation}
\sigma(\Gamma)=\int_0^\tau dt\,\varphi(x)\circ\dot{x},
\end{equation}
where $\varphi(x):=j^{\rm ss}/D(x)p^{\rm ss}(x)$ and $\circ$ denotes the Stratonovich product.

A generic time-integrated current takes the form of $\phi(\Gamma)=\int_0^\tau dt\,f(x)\circ\dot{x}$, where $f(x)$ is the projection function.
The entropy production current corresponds to the case of $f(x)=\varphi(x)$.
We consider a finite set of basis currents defined as $\phi_i(\Gamma)=\int_0^\tau dt\,f_i(x)\circ\dot{x}$, where $f_i(x)$ is the basis function.
We seek basis functions that have a rich representation, i.e., where an arbitrary function $f(x)$ can be well approximated by a linear combination of $\{f_i(x)\}_{i=1}^n$ for a certain region of $x$.
For example, $\{f_i(x)\}$ can be trigonometric functions of the Fourier basis, $\{\sin(ix),\cos(ix)\}$, or Gaussian radial basis function kernels, $\exp\bras{-(x-x_i)^2/2\vartheta_i^2}$, where $x_i$ and $\vartheta_i$ are the center and the bandwidth of the kernel, respectively.
As other choices, $\{f_i(x)\}$ can be orthogonal polynomials such as Legendre or Chebyshev polynomials \cite{Quarteroni.2007}.
In all examples, we employ trigonometric functions and Gaussian kernels and determine that they provide excellent approximations.
Theoretically, increasing the number of basis currents will enhance the representation ability.
However, as shown later, the truncation of $n$ to some order is sufficient to obtain good estimates.

Once the basis functions $\{f_i(x)\}$ are determined, the corresponding set of basis currents is $\mca{C}=\{\phi_i\}$, where $\phi_i(\Gamma)=\int_0^\tau dt\,f_i(x)\circ\dot{x}$.
Using the coefficient vector, which is calculated via the means and covariances of basis currents using Eq.~\eqref{eq:opt.c.lambda}, one can construct the optimal current as $\phi_{\rm opt}(\Gamma)=\int_0^\tau dt\,f_{\rm opt}(x)\circ\dot{x}$, where $f_{\rm opt}(x)=\sum_ic_if_i(x)$.

\section{Applications}
In this section, we apply the proposed method to three systems: the four-state Markov jump process, the periodically driven particle, and the bead-spring model.
For each system, we run a simulation and obtain a single trajectory of length $\mca{T}_{\rm obs}$, from which we estimate the entropy production rate.
Specifically, for Langevin systems, we use the Euler method to numerically solve the system dynamics with a time step of $\Delta t=10^{-4}$.
To examine the stability of the proposed method, we independently perform $20$ estimations and calculate the mean and standard deviation of the estimates for each parameter setting.

\subsection{Four-state Markov jump process}

\begin{figure}[t]
\centering
\includegraphics[width=\linewidth]{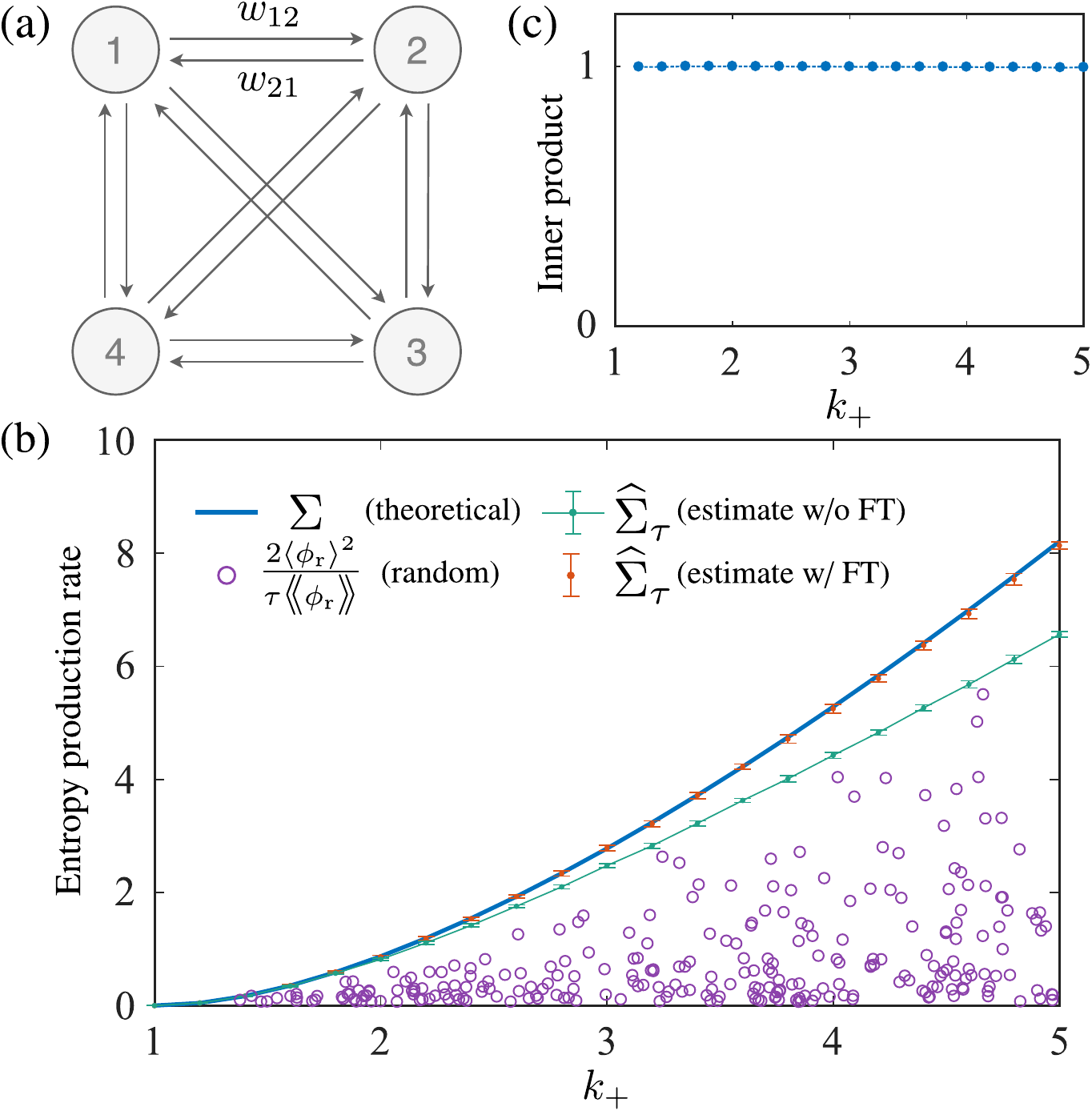}
\protect\caption{(a) Schematic diagram of the four-state Markov jump process whose states are fully connected. (b) Estimation of the entropy production rate. The blue solid line represents the actual entropy production rate, while the green solid line with dots represents its estimated tightest lower bound. The error bar depicts the standard deviation of the estimated values. The violet circles denote the lower bound on the basis of individual random currents. The orange dots with error bars represent the estimated values by combining the proposed method with the fluctuation theorem. When the entropy production is the optimal current, $\Sigma$ can be accurately estimated with the help of the fluctuation theorem. (c) Cosine similarities between the coefficients of the computed optimal current and those of the entropy production current. As shown, all inner products are close to $1$ for all $k_{+}\in[1,5]$, which empirically indicates that the entropy production is the optimal current. The parameter $k_{+}$ is varied while the remaining parameters are fixed as $k_{-}=1,\mca{T}_{\rm obs}=10^4$, and $\tau=10^{-2}$.}\label{fig:4stateMarkov}
\end{figure}

We consider the four-state Markov jump process \cite{Gingrich.2016.PRL}, whose transition rates are given as follows:
\begin{equation}
[w_{yz}]=\begin{bmatrix}
0 & k_{+} & k_{+} & k_{-}\\
k_{-} & 0 & k_{+} & k_{+}\\
k_{-} & k_{-} & 0 & k_{+}\\
k_{+} & k_{-} & k_{-} & 0
\end{bmatrix},
\end{equation}
where $k_{+}$ and $k_{-}$ are positive parameters [see Fig.~\ref{fig:4stateMarkov}(a) for illustration].
When $k_{+}=k_{-}$, the system relaxes to an equilibrium after a long period of time.
By solving the master equation, one can readily obtain the steady-state distribution
\begin{equation}
[p^{\rm ss}(y)]=\frac{1}{10 k_{-}^2+12 k_{-} k_{+}+10 k_{+}^2}\begin{bmatrix}
4 k_{-}^2+2 k_{-} k_{+}+2 k_{+}^2\\
3 k_{-}^2+4 k_{-} k_{+}+k_{+}^2\\
k_{-}^2+4 k_{-} k_{+}+3 k_{+}^2\\
2 k_{-}^2+2 k_{-} k_{+}+4 k_{+}^2
\end{bmatrix}.
\end{equation}
Using $[p^{\rm ss}(y)]$, the entropy production rate can be immediately calculated
\begin{equation}
\Sigma=\sum_{y<z}\bras{p^{\rm ss}(y)w_{yz}-p^{\rm ss}(z)w_{zy}}\ln\frac{p^{\rm ss}(y)w_{yz}}{p^{\rm ss}(z)w_{zy}}.
\end{equation}

We apply the proposed method to estimate the tightest lower bound on the entropy production rate from the single trajectory $\Gamma$ of length $\mca{T}_{\rm obs}=10^4$, which is obtained from the simulation using the Gillespie algorithm \cite{Gillespie.1977.JPC}.
The value of $k_{-}$ is fixed to $1$, while $k_{+}$ is varied in the range of $[1,5]$.
We illustrate the estimated results in Fig.~\ref{fig:4stateMarkov}(b).
As can be seen, the estimated lower bound on $\Sigma$ is tight and coincides with the actual entropy production rate when the system is close to equilibrium, i.e., when $k_{+}/k_{-}\to 1$.
When $k_{+}/k_{-}\gg 1$, the gap between the estimated value and the actual value increases, which implies that TUR cannot be saturated in this regime even with the short-time limit. 
We also generate random coefficients $\gamma_{yz}\in[-1,1]$ and form random currents $\phi_{\rm r}=\sum_{y<z}\gamma_{yz}\phi_{yz}$.
We evaluate the fluctuation of each random current, $2\avg{\phi_{\rm r}}^2/\tau\Var{\phi_{\rm r}}$ (which is a lower bound on $\Sigma$) and plot the result in Fig.~\ref{fig:4stateMarkov}(b).
Clearly, the estimated lower bound $\widehat{\Sigma}_\tau$, which is based on the optimal current, is always better than the one that is based on each individual random current.

We investigate the form of the computed optimal current by measuring the distance between the coefficients of $\phi_{\rm opt}$ and those of the entropy production $\sigma$.
Specifically, we normalize the coefficient vectors, $\widehat{\bm{\gamma}}=\bm{\gamma}/\|\bm{\gamma}\|_2$, and calculate their inner product.
Here, $\|\cdot\|_2$ denotes the Euclidean norm.
We vary $k_{+}$ and plot the cosine similarities in Fig.~\ref{fig:4stateMarkov}(c).
The cosine similarity between $\bm{\gamma}_1$ and $\bm{\gamma}_2$ is defined as $\widehat{\bm{\gamma}}_1\cdot\widehat{\bm{\gamma}}_2$, where $\cdot$ denotes the inner product of two vectors.
Interestingly, the inner products are always approximately equal to $1$, which implies that $\phi_{\rm opt}$ is identical to the current of entropy production (by ignoring the scaling factor).
Thus, $\sigma=\alpha\phi_{\rm opt}$, where $\alpha\in\mbb{R}$ is the unknown scaling factor.
Therefore, we use the fluctuation theorem to further estimate the entropy production rate, as demonstrated in the previous section (i.e., not the lower bound but the exact value of $\Sigma$).
We solve the equation $\Psi(\alpha)=\mca{N}_\Gamma$ using the Newton--Raphson method to find the nontrivial solution $\alpha\neq 0$.
Then, we estimate the entropy production rate as $\widehat{\Sigma}_\tau=\alpha\avg{\phi_{\rm opt}}/\tau$.
We plot the estimated results in Fig.~\ref{fig:4stateMarkov}(b).
As illustrated, the method in combination with the fluctuation theorem produces accurate estimates even when the system is far from equilibrium.

\subsection{Periodically driven particle}

\begin{figure}[t]
\centering
\includegraphics[width=\linewidth]{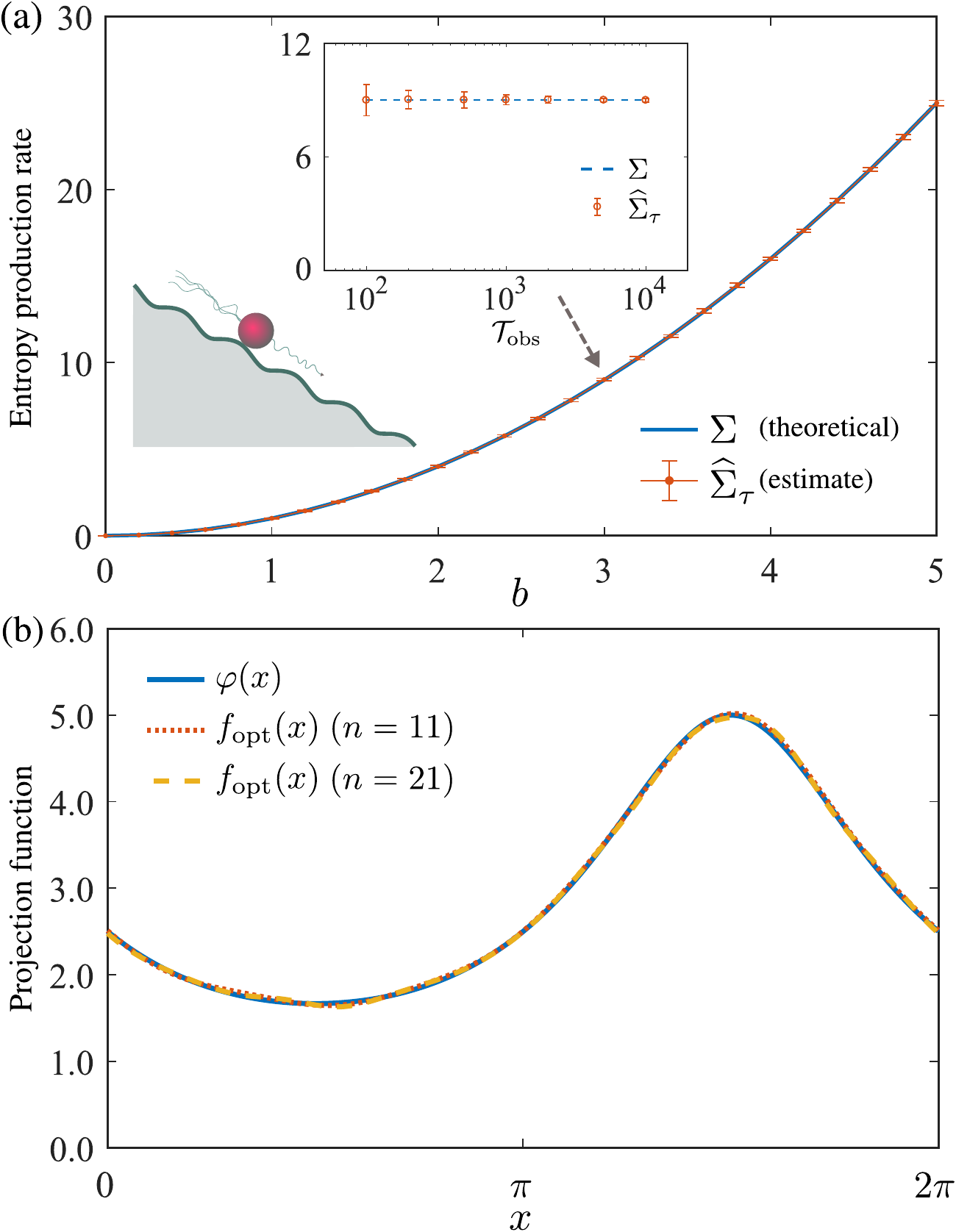}
\protect\caption{(a) Estimation of the entropy production rate $\Sigma$ in the periodically driven particle system. The blue solid line depicts the theoretical entropy production rate. The orange solid line with dots represents the mean of the estimates of $\Sigma$, and the error bars represent the standard errors. The inset shows how the estimation results are affected when the length $\mca{T}_{\rm obs}$ is changed (at $b=3$). (b) Comparison between the projection function of the computed optimal current $f_{\rm opt}(x)$ and that of the entropy production current (which is theoretically the optimal one) $\varphi(x)$ in two cases: $n=11$ and $n=21$ basis currents, when $b=5$. The solid, dotted, and dashed lines represent $\varphi(x)$, $f_{\rm opt}(x)~(n=11)$, and $f_{\rm opt}(x)~(n=21)$, respectively. The result shows that the optimal current is well approximated in both cases. The parameters are fixed as $a=2,\mca{T}_{\rm obs}=10^4$, and $\tau=10^{-2}$.}\label{fig:PeriodicParticle}
\end{figure}

Next, we consider a Brownian particle that circulates on a ring with a circumference of $2\pi$ \cite{Hasegawa.2019.PRE}, and its dynamics are governed by the Langevin equation with $F(x)=[a+\sin(x)][b+\cos(x)]$ and $D(x)=[a+\sin(x)]^2$, where $a>1$ and $b\ge 0$ are the parameters.
The effective potential is
\begin{equation}
U(x)=-\frac{1}{2}[a+\sin(x)]^2-b[ax-\cos(x)],
\end{equation}
which is illustrated in Fig.~\ref{fig:PeriodicParticle}(a).
Although the system is nonlinear, the steady-state distribution can be analytically calculated
\begin{equation}
p^{\rm ss}(x)=\frac{c}{a+\sin(x)},
\end{equation}
where $c>0$ is the normalization constant such that $\int_0^{2\pi}dx\,p^{\rm ss}(x)=1$.
The entropy production rate is given by
\begin{equation}
\Sigma=\int_0^{2\pi}dx\frac{(j^{\rm ss})^2}{D(x)p^{\rm ss}(x)}=b^2,
\end{equation}
where $j^{\rm ss}=bc$ is the probability current.
It has been shown that the equality of TUR can be exactly attained with the current of entropy production \cite{Hasegawa.2019.PRE}
\begin{equation}
\sigma(\Gamma)=\int_0^\tau dt\,\varphi(x)\circ\dot{x}
\end{equation}
for arbitrary observation time $\tau$, where $\varphi(x)=b/[a+\sin(x)]$.

To compute the optimal current, we employ basis currents with the following projection functions:
\begin{equation}
f_i(x)=\begin{cases}
1+\cos(mx), & \text{if}~i=2m+1,\\
1+\sin(mx), & \text{if}~i=2m,
\end{cases}
\end{equation}
for $i=1,\dots,n$.
Here, $1$ is added to each projection function to avoid vanishing currents.
We fix $a=2$ and vary $b$ in the range of $[0,5]$.
For each parameter setting, we use $n=21$ basis currents to approximate the optimal current.
We plot the mean and the standard error of the estimated results over $20$ independent trajectories in Fig.~\ref{fig:PeriodicParticle}(a).
It is observed that the estimated value $\widehat{\Sigma}_\tau$ and the actual entropy production rate $\Sigma$ agree well for all $b$.
The errors are always small even when $\Sigma$ increases, which confirms the stability of our method.
We also investigate the effect of the length of the trajectory on the estimation result.
We vary the value of $\mca{T}_{\rm obs}$ in the range of $[10^2,10^4]$ and plot the results in the inset of Fig.~\ref{fig:PeriodicParticle}(a).
As illustrated, the estimator is unbiased for all finite lengths of the trajectory.
The mean of the estimated values is always approximately equal to the actual entropy production rate, even when the trajectory is not long.
Compared to when $\mca{T}_{\rm obs}$ is large, the standard error tends to increase when $\mca{T}_{\rm obs}$ is small.
This occurs due to the limited length of the trajectory (i.e., there are statistical errors in the calculation of moments of basis currents).

We define
\begin{equation}
f_{\rm opt}(x):=\sum_{i=1}^{n}c_if_i(x),
\end{equation}
which is the projection function of the computed optimal current.
We plot $f_{\rm opt}(x)$ and $\varphi(x)$ in Fig.~\ref{fig:PeriodicParticle}(b) to examine whether the computed function is close to the optimal one or not.
We consider two cases: using $n=11$ and $n=21$ basis currents.
We find that $f_{\rm opt}(x)$ and $\varphi(x)$ are almost identical in both cases, which implies that the theoretically optimal current is approximated well by our method, even when using a small number of basis currents, $n=11$.
\begin{figure*}[t]
	\centering
	\includegraphics[width=\linewidth]{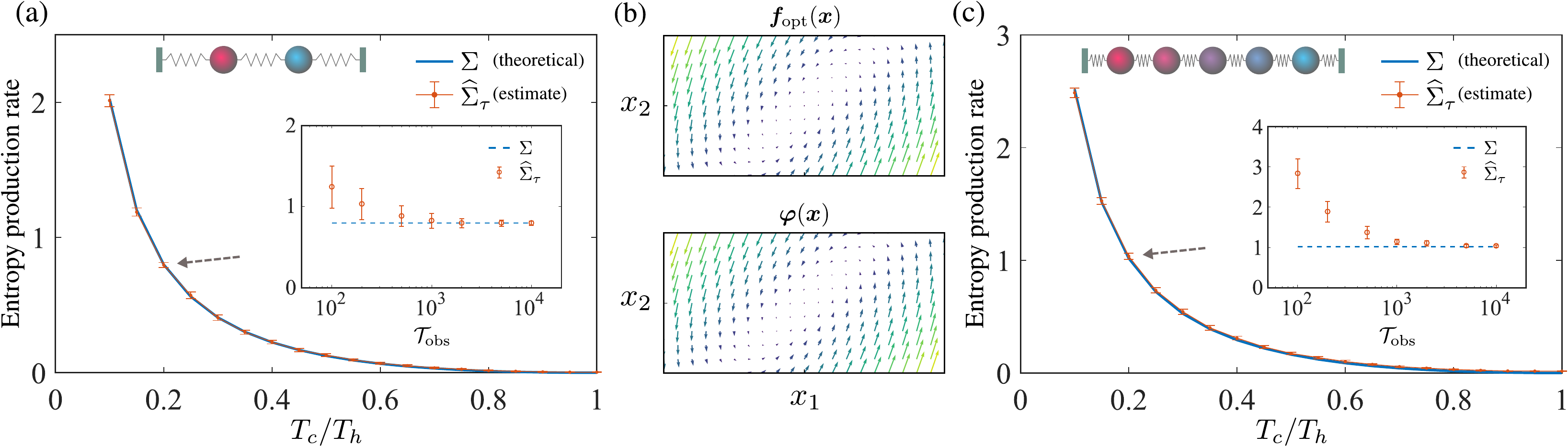}
	\protect\caption{(a) Estimation of the entropy production rate in the two-bead system. The blue solid line represents the actual entropy production rate $\Sigma$. The orange solid line with dots depicts the estimated values $\widehat{\Sigma}_\tau$, while the error bars indicate standard deviations. Blue and orange solid lines almost overlap, which implies that $\Sigma$ is accurately estimated. The inset shows the performance of the estimator when the length of the trajectory is changed. With an increase in $\mca{T}_{\rm obs}$, the estimated value converges to the exact value of $\Sigma$ with high stability. (b) Comparison between the projection function of the computed optimal current $\bm{f}_{\rm opt}(\bm{x})$ (top panel) and that of the entropy production current $\bm{\varphi}(\bm{x})$ (bottom panel). Two vector fields show the same behavior in both direction and magnitude, which empirically verifies that the optimal current is approximated well. (c) Estimation of the entropy production rate in the five-bead system. The blue and orange solid lines represent the actual entropy production rate $\Sigma$ and the estimate $\widehat{\Sigma}_\tau$, respectively. The error bars depict the standard deviations of the estimated values. The inset shows the estimation performance when the length $\mca{T}_{\rm obs}$ of the trajectory is varied. Parameter $T_h$ is varied while the remaining parameters are fixed as $k=1$ (two beads) and $4$ (five beads), $T_c=10,\mca{T}_{\rm obs}=10^4$, and $\tau=10^{-2}$.}\label{fig:BeadSpring}
\end{figure*}

\subsection{Bead-spring model}

Finally, we consider a nonequilibrium system that consists of $N$ beads that are coupled in one dimension \cite{Li.2019.NC}.
Each bead is in contact with a thermal reservoir at different temperature.
The dynamics of the system are described by the multivariate Langevin equation
\begin{equation}
\dot{\bm{x}}=\msf{A}\bm{x}+\sqrt{2\msf{D}}\bm{\xi},
\end{equation}
where $\bm{x}=[x_1,\dots,x_N]^\top$ denotes the positions of the beads and $\msf{A}\in\mbb{R}^{N\times N}$ and $\msf{D}\in\mbb{R}^{N\times N}$ are the drift and diffusion terms, respectively.
Note that $\msf{D}={\rm diag}(D_1,\dots,D_N)$ is a diagonal matrix, and the noises that affect each bead are uncorrelated.
Because the forces are linear, the steady-state distribution is Gaussian, 
\begin{equation}
p^{\rm ss}(\bm{x})=\frac{1}{\sqrt{(2\pi)^N|\msf{C}|}}\exp\bra{-\frac{1}{2}\bm{x}^\top\msf{C}^{-1}\bm{x}}.
\end{equation}
Here, $\msf{C}$ is the covariance matrix of $\bm{x}$, given by $\msf{C}_{ij}=\avg{x_ix_j}-\avg{x_i}\avg{x_j}$.
The probability current in the Fokker--Planck equation is
\begin{equation}
\bm{j}^{\rm ss}(\bm{x})=\bra{\msf{A}\bm{x}-\msf{D}\bm{\nabla}_{\bm{x}}}p^{\rm ss}(\bm{x})=(\msf{A}+\msf{D}\msf{C}^{-1})\bm{x}p^{\rm ss}(\bm{x}).
\end{equation}
The current of the stochastic entropy production reads $\sigma(\Gamma)=\int dt\,\bm{\varphi}(\bm{x})^\top\circ\dot{\bm{x}}$, where
\begin{equation}
\bm{\varphi}(\bm{x})=\msf{D}^{-1}\bm{j}^{\rm ss}(\bm{x})/p^{\rm ss}(\bm{x})=(\msf{D}^{-1}\msf{A}+\msf{C}^{-1})\bm{x}.
\end{equation}
Then, the entropy production rate is analytically obtained
\begin{equation}
\Sigma=\int d\bm{x}\,\bm{j}^{\rm ss}(\bm{x})^\top\bm{\varphi}(\bm{x})={\rm Tr}\bras{\msf{D}^{-1}\msf{A}\msf{C}\msf{A}^\top-\msf{C}^{-1}\msf{D}},\label{eq:ent.beads}
\end{equation}
where ${\rm Tr}[\cdot]$ is the trace operator that calculates the sum of elements on the main diagonal.

First, we consider the case of $N=2$ beads with the drift and diffusion terms given by
\begin{equation}
\msf{A}=\begin{bmatrix}
-2k & k\\
k & -2k
\end{bmatrix},
~\msf{D}=\begin{bmatrix}
T_h & 0\\
0 & T_c
\end{bmatrix}.
\end{equation}
Here, $k>0$ is the stiffness of the springs, and $T_h\ge T_c>0$ are the temperatures of the thermal reservoirs that are coupled to each bead.
From Eq.~\eqref{eq:ent.beads}, the entropy production rate can be analytically calculated
\begin{equation}
\Sigma=\frac{k(T_h-T_c)^2}{4T_hT_c}.
\end{equation}
\begin{figure*}[t]
\centering
\includegraphics[width=\linewidth]{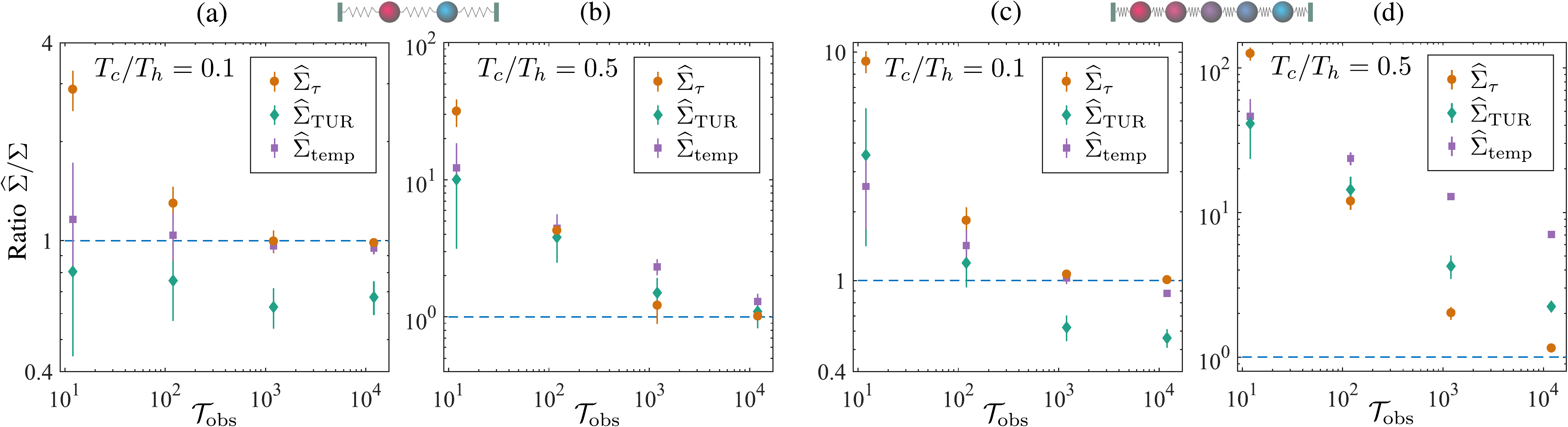}
\protect\caption{Performance of the different estimators. We compare our estimator herein, $\widehat{\Sigma}_\tau$, with the two estimators used in Ref.~\cite{Li.2019.NC}, $\widehat{\Sigma}_{\rm TUR}$ and $\widehat{\Sigma}_{\rm temp}$. The mean and standard deviation of each ratio $\widehat{\Sigma}/\Sigma$ are calculated using ten independent estimations via the two-bead model with (a) $T_c/T_h=0.1$ and (b) $T_c/T_h=0.5$, and via the five-bead model with (c) $T_c/T_h=0.1$ and (d) $T_c/T_h=0.5$. The results of the estimators $\widehat{\Sigma}_\tau$, $\widehat{\Sigma}_{\rm TUR}$, and $\widehat{\Sigma}_{\rm temp}$ are depicted using circles, diamonds, and squares, respectively. The dashed line represents the actual ratio, which equals $1$. Our estimator shows the best convergence and always provides accurate estimates when $\mca{T}_{\rm obs}$ is sufficiently long. Notably, for the five-bead model with $T_c/T_h=0.5$, estimators $\widehat{\Sigma}_{\rm TUR}$ and $\widehat{\Sigma}_{\rm temp}$ show slow convergence, while $\widehat{\Sigma}_\tau$ rapidly converges to the actual entropy production rate. The length $\mca{T}_{\rm obs}$ of the observed trajectory is varied, while the remaining parameters are fixed to $k=1$ (two beads) and $\approx 3.215$ (five beads), $T_c=25$, $\Delta t=10^{-3}$, and $\tau=10^{-2}$.}\label{fig:BeadSpringComp}
\end{figure*}

We use $m^2$ Gaussian kernels to approximate the optimal current.
Specifically, for each $i=1,\dots,m^2$, we define
\begin{equation}
f_i(\bm{x})=\exp\bras{-\frac{(\bm{x}-\bm{x}_i)^\top\msf{B}^{-1}(\bm{x}-\bm{x}_i)}{2}},
\end{equation}
where $\bm{x}_i$ is the kernel center and $\msf{B}$ is the kernel bandwidth.
From the given trajectory, we calculate $\overline{\bm{x}}=[\overline{x}_1,\overline{x}_2]^\top$, where $\overline{x}_\nu:=10+\max_{t}\brab{|x_\nu(t)|}$.
Then, $\bm{x}_i$ and $\msf{B}$ are determined as follows:
\begin{align}
\bm{x}_i&=\begin{bmatrix}
\bra{0.5+(i-1)\% m}\Delta x_1-\overline{x}_1\\
\bra{0.5+\floor{(i-1)/m}}\Delta x_2-\overline{x}_2
\end{bmatrix},\label{eq:kernel.center}\\
\msf{B}&=\begin{bmatrix}
\Delta x_1^2 & 0\\
0 & \Delta x_2^2
\end{bmatrix},
\end{align}
where $\Delta x_\nu=2\overline{x}_\nu/m~(\nu=1,2)$, $\%$ denotes the remainder of the Euclidean division, and $\floor{\cdot}$ denotes the floor function.
Equation \eqref{eq:kernel.center} indicates that the kernel centers are uniformly sampled over the region of interest, $[-\overline{x}_1,\overline{x}_1]\times[-\overline{x}_2,\overline{x}_2]$.
The optimal current is approximated using $n=2m^2$ basis currents as
\begin{align}
\phi_{\rm opt}(\Gamma)&=\int dt\sum_{i=1}^{m^2}\bras{c_{i,1}f_i(\bm{x})\circ\dot{x}_1+c_{i,2}f_i(\bm{x})\circ\dot{x}_2},\nonumber\\
&=\int dt\,\bm{f}_{\rm opt}(\bm{x})^\top\circ\dot{\bm{x}},
\end{align}
where $\bm{c}=[c_{1,1},\dots,c_{m^2,1},c_{1,2},\dots,c_{m^2,2}]$ is the coefficient vector and $\bm{f}_{\rm opt}(\bm{x}):=[\sum_{i}c_{i,1}f_i(\bm{x}),\sum_{i}c_{i,2}f_i(\bm{x})]^\top$.

We vary the temperature ratio $T_c/T_h$ in the range of $[0.1,1]$ and test the effectiveness of our method using $n=50$ basis currents (i.e., $m=5$).
For each parameter setting, we collect a trajectory of length $\mca{T}_{\rm obs}=10^{4}$, from which we estimate the entropy production rate.
We independently perform 20 estimations and obtain the mean and standard error of the estimated values.
As illustrated in Fig.~\ref{fig:BeadSpring}(a), on average, the estimator always produces an exact estimate of the entropy production rate, even when the system is far from equilibrium.
The inset shows the performance of the estimator when the length of the trajectory is changed.
Although the estimated values are biased for finite lengths, they converge to the exact values when $\mca{T}_{\rm obs}$ is increased.
In addition, the standard errors also decrease when the length $\mca{T}_{\rm obs}$ is sufficiently long.

We investigate whether the projection function of the computed optimal current $\bm{f}_{\rm opt}(\bm{x})$ is close to that of the entropy production current $\bm{\varphi}(\bm{x})$.
We plot $\bm{f}_{\rm opt}(\bm{x})$ and $\bm{\varphi}(\bm{x})$ as vector fields in Fig.~\ref{fig:BeadSpring}(b).
It is observed that these vector fields are in excellent agreement in both direction and magnitude.
This implies that $\sigma(\Gamma)$ (which is the theoretically optimal current in the short-time limit) is well approximated by the linear combination of the constructed basis currents.

Next, we consider a five-bead system, whose drift and diffusion terms are
\begin{align}
\msf{A}&=\begin{bmatrix}
-2k & k & 0 & 0 & 0\\
k & -2k & k & 0 & 0\\
0 & k & -2k & k & 0\\
0 & 0 & k & -2k & k\\
0 & 0 & 0 & k & -2k
\end{bmatrix},\\
\msf{D}&=\frac{1}{4}\begin{bmatrix}
4T_h & 0 & 0 & 0 & 0\\
0 & 3T_h+T_c & 0 & 0 & 0\\
0 & 0 & 2T_h+2T_c & 0 & 0\\
0 & 0 & 0 & T_h+3T_c & 0\\
0 & 0 & 0 & 0 & 4T_c
\end{bmatrix}.
\end{align}
For this system, the entropy production rate is equal to 
\begin{equation}
\Sigma=\frac{k(T_h-T_c)^2(111T_h^2+430T_hT_c+111T_c^2)}{495T_hT_c(3T_h+T_c)(T_h+3T_c)}.
\end{equation}
Again, we employ Gaussian kernels, whose centers and bandwidth are analogously determined as in the two-bead case.
We use $n=160$ basis currents to approximate the optimal current and plot the estimated results in Fig.~\ref{fig:BeadSpring}(c).
As shown, the estimator is unbiased for all temperature ratios $T_c/T_h$ even when the dynamics are strongly driven from equilibrium.
The inset in Fig.~\ref{fig:BeadSpring}(c) illustrates the statistics of the estimated values when the length $\mca{T}_{\rm obs}$ is changed.
The estimator is biased for small $\mca{T}_{\rm obs}$ but rapidly converges to the exact value when $\mca{T}_{\rm obs}$ is increased, which is analogous to the two-bead case.

In the end, we compare the performance of our estimator $\widehat{\Sigma}_\tau$ with that of the two estimators proposed in Ref.~\cite{Li.2019.NC}, $\widehat{\Sigma}_{\rm TUR}$ and $\widehat{\Sigma}_{\rm temp}$.
Herein, we will briefly describe these two estimators (see Ref.~\cite{Li.2019.NC} for details).
The thermodynamic force of the entropy production is estimated as $\widehat{\bm{\varphi}}(\bm{x})=\msf{D}^{-1}\widehat{\bm{j}}^{\rm ss}(\bm{x})/\widehat{p}^{\rm ss}(\bm{x})$, where $\widehat{\bm{j}}^{\rm ss}(\bm{x})$ and $\widehat{p}^{\rm ss}(\bm{x})$ are estimators of $\bm{j}^{\rm ss}(\bm{x})$ and $p^{\rm ss}(\bm{x})$, respectively.
Subsequently, $\widehat{\Sigma}_{\rm TUR}$ estimates the lower bound of the entropy production rate by utilizing the TUR with the current $\int dt\,\widehat{\bm{\varphi}}(\bm{x})^\top\circ\dot{\bm{x}}$.
On the other hand, $\widehat{\Sigma}_{\rm temp}$ directly estimates the entropy production rate via its temporal average, $\widehat{\Sigma}_{\rm temp}=\mca{T}_{\rm obs}^{-1}\int_0^{\mca{T}_{\rm obs}}dt\,\widehat{\bm{\varphi}}(\bm{x})^\top\circ\dot{\bm{x}}$.
It is worth noting that these two estimators require knowledge of the diffusion matrix $\msf{D}$, while our estimator does not rely on such information.

To evaluate the performance of the estimators, we vary the trajectory length $\mca{T}_{\rm obs}=1.2\times 10^{l}\,(1\le l\le 4)$ and focus on the convergence of each estimator.
We examine two temperature ratios, $T_c/T_h=0.1$ and $T_c/T_h=0.5$, using both the two and five-bead models.
The parameter values and experimental settings are the same as used in Ref.~\cite{Li.2019.NC}.
We calculate the mean and standard deviation of the ratio $\widehat{\Sigma}/\Sigma$ using ten independent estimations and plot them in Fig.~\ref{fig:BeadSpringComp}.
As illustrated, our estimator shows the best convergence in all cases.
When the trajectory length $\mca{T}_{\rm obs}$ is short, $\widehat{\Sigma}_\tau$ is prone to overestimating the actual entropy production rate because the trajectory does not provide sufficient information to accurately calculate the mean and variance of each basis current.
However, when $\mca{T}_{\rm obs}$ is sufficiently long, $\widehat{\Sigma}_\tau$ always obtains accurate estimates.
Notably, for the five-bead model with $T_c/T_h=0.5$, estimators $\widehat{\Sigma}_{\rm TUR}$ and $\widehat{\Sigma}_{\rm temp}$ slowly converge and return inaccurate estimates even when $\mca{T}_{\rm obs}$ is long.
In contrast, our estimator rapidly converges to the actual entropy production rate and provides the best estimate.

\section{Conclusion and Discussion}

In summary, a method for estimating entropy production based on the TUR was proposed.
Three examples, including Markov jump processes and Langevin dynamics, were studied to illustrate the effectiveness of the proposed method.
It was shown that the entropy production rate can be accurately estimated for Langevin dynamics using the short-time limit.
The results demonstrate that the estimates are significantly consistent with the theoretical entropy production rates, even when the system is far from equilibrium.
The proposed method always effectively performs, regardless of whether the noise is additive or multiplicative.
Further, it was empirically confirmed that the optimal current, which is proportional to the entropy production in the short-time limit, can be successfully approximated by the linear combination of predetermined basis currents.
Thus, the entropy production current can be accurately inferred by integrating our method with the fluctuation theorem.
Namely, one can infer not only the average of entropy production but also its probability distribution.
For Markov jump processes, our method provides the tightest lower bound for the entropy production rate.
If the condition that the entropy production current is the optimal one is given, then an exact estimate can be obtained through the combination with the integral fluctuation theorem.

From a practical perspective, the proposed algorithm can be easily implemented and is computationally efficient (i.e., all numerical computations can be performed in parallel).
The Monte Carlo sampling utilized in Ref.~\cite{Li.2019.NC} suffers from a local optimum when the dynamics are strongly driven; thus, it can be replaced by our method, which always produces a global optimum.
Unlike in Ref.~\cite{Manikandan.2019.arxiv}, where the details of underlying dynamics (e.g., the system entropy, heat, and work) are required to form the optimal current, the proposed method does not require such prior knowledge of the dynamics.

We discuss some possible future research directions.
This study focused on estimating entropy production; however, the proposed method should also apply to the estimation of the Fisher information, which is lower bounded by means and covariances of multiple observables \cite{Ito.2018.arxiv}.
Moreover, it is of interest to test our method with the experimental data.
For example, one can estimate the dissipation cost in the motor protein ${\rm F}_1$-adenosine triphosphatase \cite{Hayashi.2010.PRL} from the trajectory of the rotational angles, whose dynamics are governed by the Langevin equation.
Along with studies of applications, further research on theoretical guarantees of the proposed method is desirable.
Basically, the longer the input trajectory is, the more accurate the estimate that can be obtained is.
However, a full investigation regarding the relationship between the error of the estimate and the trajectory length is beyond the scope of this study.
The development of theoretical bounds on the error with respect to the length needs to be addressed.
In addition, overcoming the curse of dimensionality in entropy production estimation remains an open problem.
Although our proposed method works well in the five-dimensional model, it is still challenging to handle genuinely high-dimensional Langevin systems.
A considerable number of basis functions may be required to obtain an accurate approximation of the optimal current, which leads to a substantial computational cost.
As an alternative solution, one can estimate with multiple sets, whose number of basis currents is limited, and assign the largest estimated value as the lower bound of the entropy production rate.
\\\\
\emph{Note added.} We recently became aware that Shun Otsubo and his collaborators had obtained similar results \cite{Otsubo.2020.arxiv}.

\section*{Acknowledgments}
This work was supported by Ministry of Education, Culture, Sports, Science and Technology (MEXT) KAKENHI Grant No.~JP19K12153.

\appendix

\section{Saturation of TUR for Langevin dynamics in the short-time limit}\label{app:Langevin}

We prove that TUR is saturated with the current of entropy production in the $\tau\to 0$ limit.
We consider a general multivariate Langevin system, whose dynamics are described by uncorrelated Ito stochastic differential equations,
\begin{equation}
\dot{x}_i=F_i(\bm{x})+\sqrt{2D_i(\bm{x})}\xi_i(t),
\end{equation}
where $\bm{x}=[x_1,\dots,x_N]^\top$ is the vector of variables.
The current of stochastic entropy production can be expanded up to the first order of $\tau$ as
\begin{equation}
\sigma(\Gamma)=\int_0^\tau dt\,\bm{\varphi}(\bm{x})^\top\circ\dot{\bm{x}}=\bm{\varphi}(\bm{x}_0)^\top(\bm{x}_\tau-\bm{x}_0)+O(\tau),\label{eq:sigma.first.order}
\end{equation}
where $\bm{\varphi}(\bm{x}):=[D_i(\bm{x})^{-1}j_i^{\rm ss}(\bm{x})/p^{\rm ss}(\bm{x})]^\top\in\mbb{R}^{N\times 1}$ and $j_i^{\rm ss}(\bm{x})=F_i(\bm{x})p^{\rm ss}(\bm{x})-\pp_{x_i}[D_i(\bm{x})p^{\rm ss}(\bm{x})]$ is the probability current.
The average of the entropy production is given by \cite{Spinney.2012.PRE}
\begin{equation}
\avg{\sigma}=\tau\int d\bm{x}\,\sum_{i=1}^N\frac{j_i^{\rm ss}(\bm{x})^2}{D_i(\bm{x})p^{\rm ss}(\bm{x})}.
\end{equation}
Using the short-time propagator \cite{Risken.1989}, the transition probability can be written as
\begin{equation}
p(\bm{x}_\tau|\bm{x}_0)=\prod_{i=1}^N\frac{1}{\sqrt{4\pi D_i(\bm{x}_0)\tau}}\exp\bra{-\frac{[x_{i,\tau}-x_{i,0}-\tau F_i(\bm{x}_0)]^2}{4D_i(\bm{x}_0)\tau}}.\label{eq:trans.prob}
\end{equation}
Here, $x_{i,0}:=x_i(0)$, $x_{i,\tau}:=x_i(\tau)$, and $p(\bm{x}_\tau|\bm{x}_0)$ denotes the conditional probability distribution that the system is in $\bm{x}_\tau$ at time $t=\tau$, given that the system is initially in $\bm{x}_0$ at time $t=0$.
Using Eqs.~\eqref{eq:sigma.first.order}--\eqref{eq:trans.prob}, the variance of entropy production can be analytically calculated as
\begin{equation}
\begin{aligned}
\Var{\sigma}&=\avg{\sigma^2}-\avg{\sigma}^2\\
&=\int d\bm{x}_0\,p^{\rm ss}(\bm{x}_0)\int d\bm{x}_\tau\,p(\bm{x}_\tau|\bm{x}_0)\\
&\times\bras{\bm{\varphi}(\bm{x}_0)^\top(\bm{x}_\tau-\bm{x}_0)+O(\tau)}^2+O(\tau^2)\\
&=2\tau\int d\bm{x}\,\sum_{i=1}^N\frac{j_i^{\rm ss}(\bm{x})^2}{D_i(\bm{x})p^{\rm ss}(\bm{x})}+O(\tau^2)\\
&=2\avg{\sigma}+O(\tau^2).
\end{aligned}
\end{equation}
Note that to obtain the third equality, means and covariances of $\bm{x}_\tau-\bm{x}_0$ are calculated by employing properties of the Gaussian distribution given in Eq.~\eqref{eq:trans.prob}.
Specifically, $\avg{x_{i,\tau}-x_{i,0}}=\tau F_i(\bm{x}_0)$ and $\avg{(x_{i,\tau}-x_{i,0})(x_{j,\tau}-x_{j,0})}=\delta_{ij}\bras{\tau^2F_i(\bm{x}_0)^2+2D_i(\bm{x}_0)\tau}$, where the average is taken over distribution $p(\bm{x}_\tau|\bm{x}_0)$ and $\bm{x}_0$ is fixed.
Subsequently, the Fano factor can be written as
\begin{equation}
\mca{F}=\frac{\Var{\sigma}}{\avg{\sigma}}=2+O(\tau).
\end{equation}
Thus, one can easily confirm that the Fano factor of entropy production converges to $2$ as $\tau\to 0$; equivalently, TUR is saturated in the short-time limit with the current of entropy production.

\section{Counterexample for the unattainability of TUR in Markov jump processes}\label{app:Markov.jump}

We show an example of Markov jump processes in which TUR is not saturated with the current of the entropy production in the short-time limit.
Explicitly, we consider a ring-type Markov chain with $N$ states, $\{1,2,\dots,N\}$.
For each $i=1,\dots,N$, a forward jump from state $i$ to state $i+1$ occurs at the rate of $k_{+}>0$, and a backward jump from state $i+1$ to state $i$ occurs at the rate of $k_{-}>0$.
Here, state $N+1$ is identical to state $1$.
There are no other transitions between nonconsecutive states.
In the short-time limit, i.e., $\tau\to 0$, the mean and variance of entropy production can be calculated as
\begin{align}
\avg{\sigma}&=\tau(k_{+}-k_{-})\ln\frac{k_{+}}{k_{-}},\\
\Var{\sigma}&=\tau(k_{+}+k_{-})\bra{\ln\frac{k_{+}}{k_{-}}}^2 + O(\tau^2).
\end{align}
Subsequently, we can obtain the Fano factor $\mca{F}$ of the entropy production
\begin{equation}
\mca{F}=\frac{\Var{\sigma}}{\avg{\sigma}}\xrightarrow{\tau\to 0}\frac{k_{+}+k_{-}}{k_{+}-k_{-}}\ln\frac{k_{+}}{k_{-}}.
\end{equation}
It is observed that $\mca{F}$ can be arbitrarily large and does not converge to $2$ in the vanishing-time limit.
Because
\begin{equation}
\ln\frac{k_{+}}{k_{-}}\ge 2\frac{k_{+}-k_{-}}{k_{+}+k_{-}},~\forall k_{+},k_{-}>0,
\end{equation}
we have $\mca{F}\ge 2$ as $\tau\to 0$.
$\mca{F}\to 2$ only when $k_{+}/k_{-}\to 1$, which means that the system is near equilibrium.
This agrees with the conclusion in previous studies \cite{Pigolotti.2017.PRL,Hasegawa.2019.PRE} that TUR is asymptotically saturated near equilibrium for the current of entropy production.


%

\end{document}